\documentclass[11pt,twoside]{osajnl}

\journal{ol} 

\setboolean{shortarticle}{true}

\usepackage{lineno}

\title{Absolute frequencies of H$^{13}$C$^{14}$N hydrogen cyanide transitions in 1,5\,$\mu$m region with saturated spectroscopy and sub-kHz scanning laser}

\author[1]{Jan Hrabina}
\author[1]{Martin Hosek}
\author[1,*]{Simon Rerucha}
\author[1]{Martin Cizek}
\author[1]{Zdenek Pilat}
\author[2]{Massimo~Zucco}
\author[1]{Josef Lazar}
\author[1]{Ondrej Cip}

\doi{This is an arXiv rendidion of an eponymous letter published in Optic Letters \url{https://doi.org/10.1364/OL.467633}}

\affil[1]{Institute of Scientific Instruments of the CAS (ISI), Kralovopolska 147, 61264 Brno, CZ}
\affil[2]{INRIM Istituto Nazionale di Ricerca Metrologica, Strada delle Cacce 91, Torino, 10135, Italy}

\affil[*]{Corresponding author: res@isibrno.cz}

\begin{abstract}
The wide span and high density of lines in its rovibrational spectrum render hydrogen cyanide a useful spectroscopic media for referencing absolute frequencies of lasers in optical communication and dimensional metrology.
We determined, for the first time to the best of our knowledge, the molecular transitions' center frequencies of the H$^{13}$C$^{14}$N isotope in the range from $1526\,$nm to $1566\,$nm with $1,3\times10^{-10}$ fractional uncertainty. We investigated the molecular transitions with a highly coherent and widely tunable scanning laser that was precisely referenced to a Hydrogen maser through an optical frequency comb.
We demonstrated an approach to stabilize the operational conditions needed to maintain the constantly low pressure of the hydrogen cyanide to carry out the saturated spectroscopy with the third-harmonic synchronous demodulation.
We demonstrated approximately a fourtyfold improvement in the line centers' resolution compared to the previous result.
\end{abstract}

\setboolean{displaycopyright}{false}

\begin{document}

\maketitle

%
\pagestyle{plain}

\section{Introduction}

\begin{figure}[htbp]
\centering
\includegraphics[width=.95\linewidth]{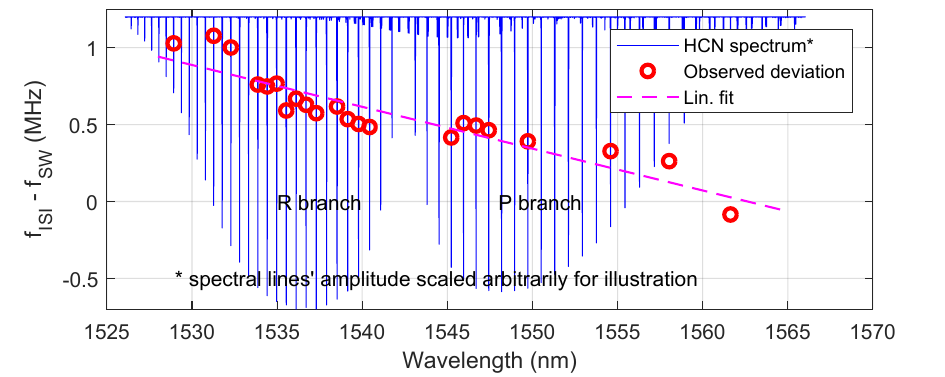}
\caption{Line centers' frequencies measured with saturated spectroscopy ($\nu_{ISI}$) compared to published data ($\nu_{SW}$) \cite{Swann2005}. The "HCN spectrum" shows hydrogen cyanide H$^{13}$C$^{14}$N 2$\nu_3$ rotational–vibrational
band spectrum obtained by scanning a tunable laser and
measuring the laser intensity transmitted through a HCN cell with an interaction length of 120 cm and 
filled to a pressure of $267\,$Pa ($2\,$Torr $\pm 10\%$); the rolling ball filter \cite{Sternberg1983} was used to remove signal background.}
\label{fig-spectrum}
\end{figure}

Two decades ago, the hydrogen cyanide (HCN) molecule emerged among spectroscopic media suitable for referencing the lasers' optical frequencies in the $1550\,$nm bands.
It brought some interesting features as an alternative to commonly used isotopes of acetylene. 
The absorption lines' wavelength band is significantly wider than that of acetylene, and it particularly fits the most widely used C-band in the wavelength division multiplexing in optical fiber communications.
The spectrum of HCN isotope H$^{13}$C$^{14}$N $2\nu_3$ band contains more than $50$ rovibrational transitions, shown in Fig \ref{fig-spectrum}, useful for spectroscopic referencing the lasers' optical frequencies, provided that the absolute frequencies of the transitions are precisely determined.

First thorough study \cite{Sasada1990} determined the frequencies of absorption line centers with the uncertainty of $\approx15\,$MHz (which translates to fraction uncertainty of $8\times10^{-8}$). 
More recent and the most exhaustive study so far \cite{Swann2005} delivered more accurate data reaching the fraction uncertainty of $5\times10^{-9}$ and added the characterization of pressure-induced shift and Doppler broadening. These works used linear spectroscopy to determine the center of absorption lines, which is, however, influenced by Doppler background.
The use of saturated absorption spectroscopy can achieve a significant improvement, such as, for example, with the acetylene where the uncertainties fall below $1\times10^{-11}$ for $^{13}C_2H_2$ isotope \cite{edwards2005atlas}. 

The activities related to the latter study, carried out by NIST\,\cite{Swann2005}, might have brought the reception and recognition of the HCN among the established spectroscopic media. 
Consequently, several commercial manufacturers started producing cells, and later on, the HCN found its way to applications beyond optical communications.

In this paper, we demonstrate the feasibility of an approach to systematically measuring the line centers' frequencies and other characteristics with (estimated) fractional uncertainty below $1,3\times10^{-10}$ level. Then we present a comparison to previously published data, where our measurement revealed a systematic deviation from previous measurements.

%

\section{Aims and challenges}

Our effort was motivated by an application in dimensional metrology, namely the optical absolute distance measurement with frequency scanning interferometry (FSI) \cite{Dale14_FSI_NPL,Hughes2017}. 
In the single-wavelength interferometry, the fractional uncertainty of laser optical frequency, i.e. the ratio of the absolute uncertainty of the optical frequency (i.e. laser linewidth) to nominal optical frequency, translates to equal uncertainty contribution in the length measurement $u_c = u_\nu / \nu$.
The effect is more  pronounced in multi-wavelength absolute measurement (such as the FSI) where the uncertainties propagates (for two wavelengths) as $u_c = (u_{\nu 1} + u_{\nu 2} ) / (\nu_1 - \nu_2)$, i.e. the ratio of sum of the linewidths to the difference of the optical frequencies. For example, with two lines at 1535 and $1550\,$nm with uncertainty of $u_{\nu 1} = u_{\nu 2} = 1\,$MHz ($5\times10^{-9}$) used for the inference of absolute distance would contribute $3\times10^{-7}$ to the uncertainty of the length measurement. This uncertainty has two major components: a) the precision of the line center resolution and b) the uncertainty of line center frequency. Our effort aimed to tackle the latter.

Surprisingly, an atlas of precisely measured centers by the saturated absorption spectroscopy of hyper-fine hydrogen cyanide transitions was still not developed, although primary time standards \cite{Li2011} and optical atomic clocks \cite{Ludlow_RMF_2015} with relative stability higher than $10^{-16}$ are available now, and optical frequency combs \cite{Fortier2019} allow the transfer of such stability to any wavelength in the visible and infrared part of the spectrum. 
To date, "many narrow lines" have been observed in \cite{Labachelerie1995}, which reported saturated spectroscopy using a cell placed inside a cavity to build up the necessary pump power but without determining the center frequencies. The insular measurement of the P27 line took place using a similar setup in \cite{Awaji1995}, and the observation of the single R7 line's Lamb dip using an HCN-filled photonic fiber was reported in \cite{Henningsen2005}.

We identified a likely reason why any complex investigation did not occur: we found it challenging to maintain the HCN in stable conditions (in terms of pressure and observed line profile intensity) over a prolonged period when the HCN pressure is kept low ($\approx$ several Pa). We attributed this effect partially to extensive adsorption and desorption of HCN molecules to gas cell walls, while we assume other effects are still present. 
 
%

\section{Methods}
\begin{figure}[htbp]
\centering
\includegraphics[width=\linewidth]{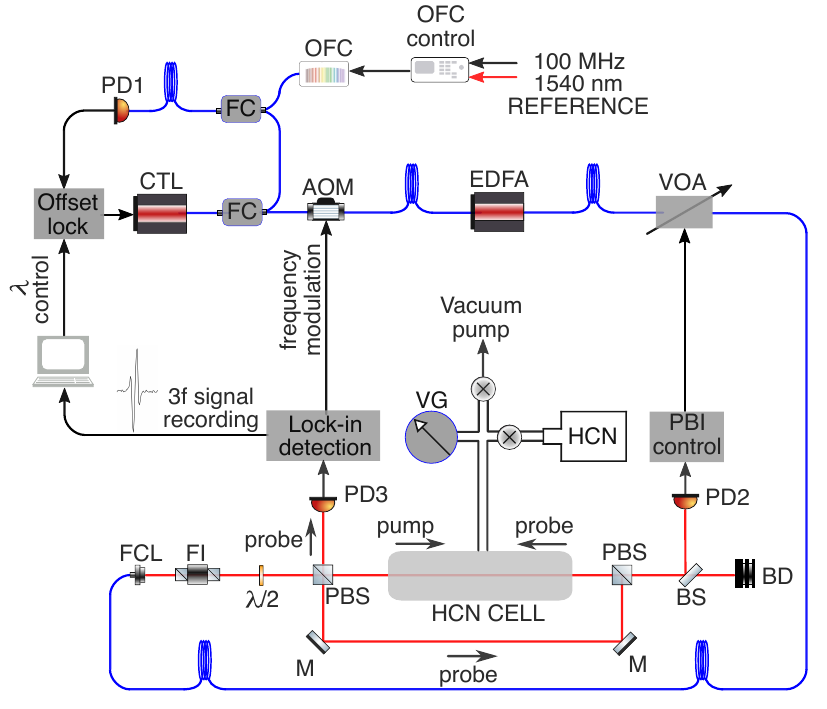}
\caption{Experimental arrangement: optical frequency comb (OFC), continuously-tuneable laser (CTL), Faraday isolator (FI), acousto-optical modulator (AOM), Er:doped fibre amplifier (EDFA), voltage-controlled optical attenuator (VOA), photodetectors (PD), polarizing beam splitters (PBS), mirrors (M), beam sampler (BS), beam dump (BD), vacuum gauge (VG), fibre couplers (FC), fibre collimator (FCL), probe beam intensity(PBI).}
\label{fig-setup}
\end{figure}

We developed an experimental apparatus, shown in Fig.~\ref{fig-setup}, which allowed us to investigate the line centers' frequencies of HCN with high precision and absolute accuracy. 
Three principal parts of the setup were a) the widely tunable laser referenced to an optical frequency comb, b) the vacuum system optimized for the HCN fine-resolution pressure control, and c) the optical arrangement for saturated spectroscopy. 
For each investigated line, we used the precise control of the laser's optical frequency to scan over the line profile, record the amplitude profile of the third-harmonic error signal, and detect its zero crossing, which corresponds to the center frequency.

%
%
\subsection{Highly coherent scanning laser}
The optical setup was powered by the continuously-tunable extended cavity diode laser (CTL) that features a mode-hop-free tuning range from $1510\,$nm to $1630\,$nm (CTL 1550, TOPTICA Photonics AG). 
CTL's optical frequency was locked to a selected tooth of the optical frequency comb (OFC) using optical mixing and in-house developed digitally controlled frequency offset lock, which incorporated a direct digital synthesizer (18-bit resolution) and a fast phase-locked loop unit ($450\,$kHz bandwidth; based on AD9956 chip).
This scheme allows us to precisely control the CTL's output frequency and also narrow the linewidth of the emission spectral profile from initial $\approx300\,$kHz down to $\approx100\,$Hz level. 
The OFC itself was optically referenced to a laser that %
was stabilized to an ultra-narrow linewidth optical cavity at $1540\,$nm. 
The repetition frequency of OFC ($250\,$MHz) was disciplined by a H-maser \cite{Cizek2022}. 
Its absolute frequency is maintained by continuous tracking to a global navigation satellite system (GNSS) time through time transfer receiver instrument \cite{panek2012}(DICOM GTR50), and non-periodic comparisons with $^{40}$Ca$^+$ ion optical clock \cite{Cip2018}.
With this arrangement, the absolute frequency of CTL was well known with $10^{-13}$ accuracy and controllable down to the $10^{-15}$ level. 

\subsection{Vacuum system with pressurized absorption cell}
The vacuum part of the setup consisted of the absorption cell, reservoir with HCN gas, vacuum gauge, and turbomolecular vacuum pump. The body of the $40\,$cm long absorption cell was made of stainless steel equipped with a vacuum flange, enabling interconnection to the vacuum system. 
The cell's optical windows were made of fused silica and glued to the cell body by vacuum-compatible glue. The amount of gas could be precisely dosed by a needle valve from the reservoir filled with H$^{13}$C$^{14}$N (Wavelength References, Inc.), and the pressure inside the cell was monitored with a capacitance vacuum gauge. 
The vacuum gauge offset was zeroed before each filling of the cell to keep the absolute scale of the gauge stable and repeatable. 

To mitigate the low-pressure effects (such as the adsorption/desorption of HCN molecules to/from the cell body), the inside of the cell was activated by perfusion with ozonized air for 90 minutes, with $\approx 50\,$mg/h of ozone. 
Subsequently, the cell was filled with a 1\% solution of chlorotrimethylsilane (Aldrich, >98\% by GC) in methanol (Penta chemicals, p.a.) and incubated for 90 minutes at room temperature. 
The solution was then discarded, and the cell was thoroughly washed with pure methanol, blown dry with filtered air, and simultaneously heated to about $50\,^\circ$C with a hot air gun.

\subsection{Saturated spectroscopy setup}
The investigation of the HCN absorption lines used a saturated spectroscopy optical setup. 
The output beam of the CTL was frequency modulated with an acousto-optical
modulator/shifter (AOM)  ($5\,$kHz modulation frequency, $3\,$MHz FM deviation)
and amplified in Er:doped fibre amplifier (EDFA) with $\approx200\,$mW output power. 
A voltage-controlled optical attenuator (VOA; V1550PA, Thorlabs) suppressed residual amplitude fluctuations.
The laser light was fed through an optical fiber collimator and Faraday isolator into the free-space part of the optical setup, where it split on polarizing beam splitters into two counter-propagating beams: pump ($\approx 120\,$mW) and probe ($\approx 10\,$mW) with diameters of $2,3\,$mm.
The probe signal from a 40 cm long refillable absorption cell was synchronously demodulated using a digital lock-in amplifier (SRS 865, Standford Research Systems) with third harmonic detection. 
We would like to note that we identified the analog modulation signal as susceptible to residual offsets from zero and zero-drifts, which could cause significant shifts in the observed line centers' frequencies.

\subsection{Measurement of the line centers' frequencies}
Before actual measurement, the vacuum system was evacuated and then filled with HCN of desired pressure.
The laser frequency was detuned from the line profile, and the offset of the lock-in amplifier was zeroed (to suppress the residual amplitude modulation). 

After preparation, the scanning sequence was commenced in the frequency range of $\pm12,5\,$MHz around the line center. 
We used $1\,$kHz steps around the central part of the profile and $5\,$kHz steps in the rest of the range at the rate of $100\,$ms/step. 
The intensity signal was digitized and recorded with auxiliary information (e.g., cell pressure).

\begin{figure}[htbp]
\centering
\includegraphics[width=\linewidth]{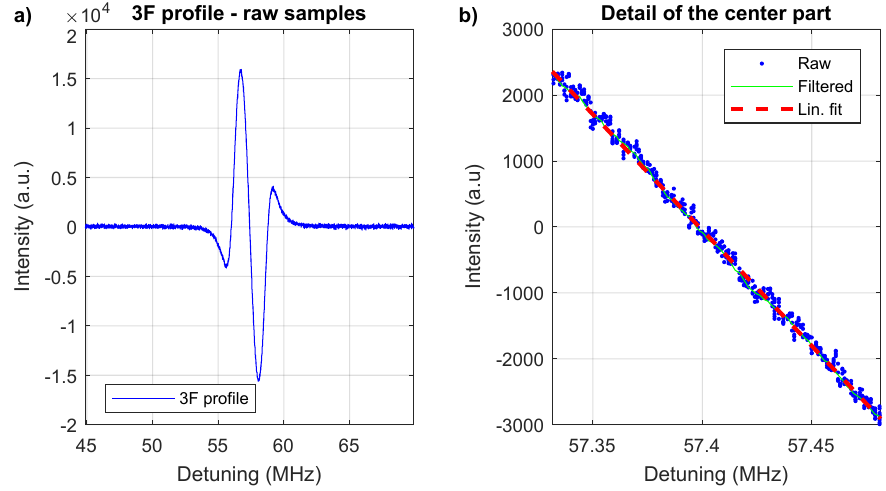}
\caption{Third-harmonic profile of the R7 line of the $2\nu_3$ band obtained using $40\,$cm long cell filled to a pressure of $1,6\,$Pa (a) and the detail of the center part with indicated linear fit (b).}
\label{fig-scanprofile}
\end{figure}

From the recorded data (typical SNR $\approx100\,$dB), shown in Fig.~\ref{fig-scanprofile}, we averaged the readings at individual points and (coarsely) removed any residual background caused by residual amplitude modulation and etalon effects in the optical part. We identified the third-harmonic profile's minimum and maximum and interpolated the points around the central part to detect the zero-crossing, which corresponds to the line's center frequency.

To investigate the measurement repeatability, we ran repeated scans of the R7 line profile (arbitrarily selected) over slightly longer than a day. 
Then we selected and measured the profiles of 17 lines (those with convenient offset from the nearest OFC tooth), ten times each. 
We carried out the measurements with pressures between $1,45\,$Pa and $1,8\,$Pa.

\section{Results and observations}
\begin{figure}[htbp]
\centering
\includegraphics[width=\linewidth]{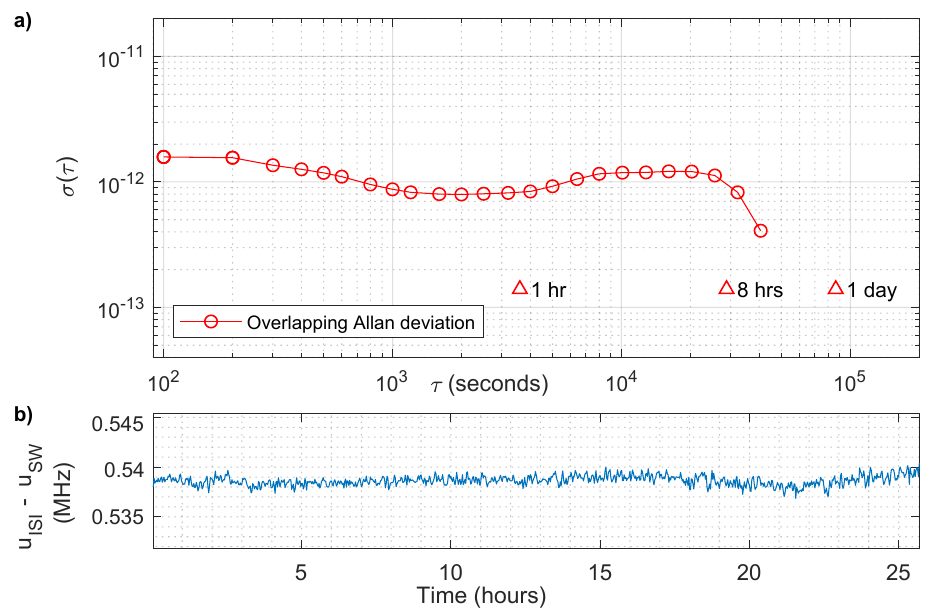}
\caption{Overlapping Allan deviations calculated from repeated scanning over the R7 absorption line (a) and the frequency recording expressed as a coincidence of measurements with saturated spectroscopy ($\nu_{ISI}$) and published data ($\nu_{SW}$)\,\cite{Swann2005}
(b).}
\label{fig-allan}
\end{figure}

The R7 line frequency recording, shown in Fig.~\ref{fig-allan} (b), reveals the standard deviation of frequency determination of $0,5\,$kHz. 
The overlapping Allan deviations, in Fig.~\ref{fig-allan} (a), reach the bottom of $8\times10^{-13}$ for $\tau \approx 1800\,$s and the entire curve fits below $1.6\times10^{-12}$, which translates to $0,3\,$kHz. The frequency recording is compensated for the pressure shifts and also for slight drifts ($0.05\,$Pa/hr) in the pressure measurement we observed during the measurement. 
Further investigation is needed here.

\begin{table}[htbp]
\centering
\caption{Center frequencies of the $2\nu_3$ band of H$^{13}$C$^{14}$N, measured in $40\,$cm cell filled to the pressure of ($1,65\pm0,1$)\,Pa and interpolated to zero pressure$^\ddagger$.}

\begin{tabular}{lccc}
\hline
Line&This work$^+$&Ref. \cite{Swann2005}*$^\alpha$&$\delta\nu$\\
    &$\nu_{ISI}\,$(MHz)&$\nu_{SW}\,$(MHz)&(MHz)\\
\hline
 R23 & 196080395.145 (25) & 196080394.150(1154) &      0.995 \\ 
 R18 & 195779623.109 (25) & 195779622.061(1023) &      1.048 \\ 
 R12 & 195379001.274 (25) & 195379000.555(1019) &      0.719 \\ 
 R11 & 195308040.399 (25) & 195308039.664(1018) &      0.734 \\ 
  R9 & 195162540.395 (25) & 195162539.755(1016) &      0.640 \\ 
  R8 & 195088004.481 (25) & 195088003.897(1016) &      0.584 \\ 
  R7 & 195012280.156 (25) & 195012279.620(1015) &      0.536 \\ 
  R5 & 194857273.056 (25) & 194857272.495(1013) &      0.561 \\ 
  R3 & 194697533.032 (25) & 194697532.583(1012) &      0.450 \\ 
  R2 & 194615892.657 (25) & 194615892.235(1011) &      0.422 \\ 
  P4 & 194011541.171 (25) & 194011540.774(1004) &      0.397 \\ 
  P5 & 193920533.536 (25) & 193920533.046(1003) &      0.491 \\ 
  P7 & 193735035.335 (25) & 193735034.897(1002) &      0.439 \\ 
 P10 & 193448113.529 (25) & 193448113.185(999) &      0.344 \\ 
 P16 & 192843268.251 (25) & 192843267.988(992) &      0.263 \\ 
 P20 & 192417282.929 (25) & 192417282.740(988) &      0.189 \\ 
 P24 & 191973284.118 (25) & 191973284.280(1106) &     -0.162 \\ 
\hline
\end{tabular}
\vskip 0.08cm
{\parskip = 0cm
  $^\ddagger$ measured pressure coefficient from \cite{Swann2005}, Tab.1 are considered.\\
  $^+$ a general estimate of the uncertainty is stated.\\
  *calculated values from \cite{Swann2005}, Tab. 4 are considered.\\
  $^\alpha$ the counter-intuitive slightly decreasing trend in uncertainties results from the fact the reference data are expressed as wavelength and with the precision of $\approx0,12\,$MHz.
   }
  \label{tab_data}
\end{table}

The line centers measured so far and extrapolated to zero pressure using the previously published pressure coefficients\cite{Swann2005}, displayed in Tab \ref{tab_data}, reveals approximately fourtyfold improvement in the uncertainty of the center frequency determination compared to \cite{Swann2005}. 
For the present data, the $25\,$kHz ($40\,$kHz) uncertainty were stated uniformly.
This qualified estimate took into account 
  uncertainty in pressure coefficients and pressure measurement($\approx10\,$kHz), 
 the observed spread of values with repeated measurement of individual lines ($\approx5\,$kHz), 
 the uncertainty of the fitting function ($\approx3\,$kHz), 
 residual zero-offset or profile background ($\approx3\,$kHz), 
 residual offsets an drifts in the harmonic detection analogue control ($\approx3\,$kHz), 
 and additional (large) safety margin for potential systematic effects in, e.g., the degree of purity of HCN gas, in-homogeneous density of the HCN gas in the cell, and other phenomena unknown in this period of the research. The uncertainty was increased for lines R2, R3 (40kHz) due to observed distortion in the profile, which we attribute to residual spectroscopic features in close vicinity of the measured line. 
Further investigation is required before the complete atlas will be issued. 
For the comparison with the previously published measurement, we used the calculated frequencies, their uncertainties, and pressure shifts from\,\cite{Swann2005}.

The results indicate a linear trend in the frequency deviations depending on the absorption line frequency, as shown in Fig.~\ref{fig-spectrum}. 
It reveals that the present data systematically deviate from previous data toward higher frequencies. 
The deviation lies within the stated uncertainty for the previous data (except for the R18); nonetheless, in applications using, for instance, the frequency resolution against multiple lines, this deviation might skew the measurement results significantly. 
We may attribute this systematical deviation to the fact that the wavelength meter used for the measurement in \cite{Swann2005} was referenced right at the $\lambda = 1560\,$nm, where both data sets better coincide.
The observed systematic deviation also might explain the HCN cell calibration issues discussed in \cite{Dale14_FSI_NPL}. 

%
\section{Conclusion}
We developed the widely tunable scanning laser system with sub-kHz linewidth and the absolute accuracy of $10^{-13}$. 
In conjunction with the optical assembly for saturated spectroscopy and vacuum system allowing for precise pressure control in the gas cell, we used it to measure the third-harmonic profiles of selected lines in the H$^{13}$C$^{14}$N $2\nu_3$ band. 

We measured $\approx30\%$ of the lines in the spectrum of interest with an associated uncertainty down to $25\,$kHz ($<1,3\times10^{-10}$ fractional uncertainty). The precision achieved represents about forty-fold improvement against previous investigations \cite{Swann2005}. 
To our best knowledge, a systematic investigation of such an amount of individual lines by saturated spectroscopy was not reported before. Compared to the previously published data, we generally confirmed the validity and correctness of the previous results. 
However, we identified a systematic deviation of the line centers' determination below the resolution of the original data.

We demonstrated the methodology allowing for further measurement of the complete spectra of interest, detailed investigation of the pressure effects, and stating associated uncertainties.  
Such an outcome would hopefully be a valuable addition to the current state-of-art in the well-known and accepted spectroscopic references for the realization of SI metre in the $1550\,$nm region. 
The refined spectroscopic data might help the hydrogen cyanide towards inclusion in the CIPM's {\it Recommended values of standard frequencies} \cite{Riehle_2018} and {\it Mise en pratique for the definition of the metre in the SI} \cite{Schodel2021}.

\begin{backmatter}
\bmsection{Funding}
 European Metrology Programme for Innovation and Research (17IND03 LaVA); Ministerstvo Školství, Mládeže a Tělovýchovy (CZ.1.05/2.1.00/01.0017, LO1212,  CZ.02.1.01/0.0/0.0/16\_026/0008460); Akademie Věd České Republiky (RVO:68081731).

\bmsection{Acknowledgments} Authors acknowledge the support from and fruitful discussion with the team from NPL, UK involved in the LaVA project (A. Lewis, B. Hughes, M. Campbell). 
Authors acknowledge the support with HCN-related calculations from our dear colleague Prof. Frédéric Du Burck from Laboratoire de Physique des Lasers, Université Sorbonne Paris Nord.
The project 17IND03 LaVA has received funding from the EMPIR programme co-financed by the Participating States and from the European Union's Horizon 2020 research and innovation. The research used infrastructure supported by MEYS CR, EC, and CAS (LO1212, CZ.1.05/2.1.00/01.0017, RVO:68081731).
\bmsection{Disclosures} The authors declare no conflicts of interest.

\bmsection{Data availability} Data underlying the results presented in this paper are not publicly available at the time of publication but may be obtained from the authors upon reasonable request.

\end{backmatter}


\begin{thebibliography}{10}
	\newcommand{\enquote}[1]{``#1''}
	
	\bibitem{Sasada1990}
	H.~Sasada and K.~Yamada, \enquote{{Calibration lines of HCN in the 1.5-$\mu$m
			region},} {\protect\JournalTitle{Applied Optics}} \textbf{29}, 3535 (1990).
	
	\bibitem{Swann2005}
	W.~C. Swann and S.~L. Gilbert, \enquote{{Line centers, pressure shift, and
			pressure broadening of 1530-1560 nm hydrogen cyanide wavelength calibration
			lines},} {\protect\JournalTitle{Journal of the Optical Society of America B}}
	\textbf{22}, 1749--1756 (2005).
	
	\bibitem{edwards2005atlas}
	C.~Edwards, H.~Margolis, G.~Barwood, S.~Lea, P.~Gill, and W.~Rowley,
	\enquote{High-accuracy frequency atlas of $\mathrm{^{13}C_{2}H_{2}}$ in the
		1.5 $\mu$m region,} {\protect\JournalTitle{Applied Physics B}} \textbf{80},
	977--983 (2005).
	
	\bibitem{Sternberg1983}
	S.~R. Sternberg, \enquote{{Biomedical Image Processing},}
	{\protect\JournalTitle{Computer}} \textbf{16}, 22--34 (1983).
	
	\bibitem{Dale14_FSI_NPL}
	J.~Dale, B.~Hughes, A.~J. Lancaster, A.~J. Lewis, A.~J.~H. Reichold, and M.~S.
	Warden, \enquote{Multi-channel absolute distance measurement system with sub
		ppm-accuracy and 20 m range using frequency scanning interferometry and gas
		absorption cells,} {\protect\JournalTitle{Opt. Express}} \textbf{22},
	24869--24893 (2014).
	
	\bibitem{Hughes2017}
	B.~Hughes, M.~A. Campbell, A.~J. Lewis, G.~M. Lazzarini, and N.~Kay,
	\enquote{{Development of a high-accuracy multi-sensor, multi-target
			coordinate metrology system using frequency scanning interferometry and
			multilateration},} {\protect\JournalTitle{Videometrics, Range Imaging, and
			Applications XIV}} \textbf{10332}, 1033202 (2017).
	
	\bibitem{Li2011}
	R.~Li, K.~Gibble, and K.~Szymaniec, \enquote{Improved accuracy of the
		{NPL}-{CsF}2 primary frequency standard: evaluation of distributed cavity
		phase and microwave lensing frequency shifts,}
	{\protect\JournalTitle{Metrologia}} \textbf{48}, 283--289 (2011).
	
	\bibitem{Ludlow_RMF_2015}
	A.~D. Ludlow, M.~M. Boyd, J.~Ye, E.~Peik, and P.~O. Schmidt, \enquote{Optical
		atomic clocks,} {\protect\JournalTitle{Reviews of Modern Physics}}
	\textbf{87}, 637--701 (2015).
	
	\bibitem{Fortier2019}
	T.~Fortier and E.~Baumann, \enquote{20 years of developments in optical
		frequency comb technology and applications,}
	{\protect\JournalTitle{Communications Physics}} \textbf{2} (2019).
	
	\bibitem{Labachelerie1995}
	M.~D. Labachelerie, K.~Nakagawa, Y.~Awaji, and M.~Ohtsu,
	\enquote{{High-frequency-stability laser at 1.5 um using Doppler-free
			molecular lines},} {\protect\JournalTitle{Optics Letters}} \textbf{20},
	572--574 (1995).
	
	\bibitem{Awaji1995}
	Y.~Awaji, K.~Nakagawa, M.~de~Labachelerie, M.~Ohtsu, and H.~Sasada,
	\enquote{Optical frequency measurement of the {H}$^\mathrm{12}${C}$^{14}${N}
		{L}amb-dip-stabilized 1.5-$\mu$m diode laser,} {\protect\JournalTitle{Opt.
			Lett.}} \textbf{20}, 2024--2026 (1995).
	
	\bibitem{Henningsen2005}
	J.~Henningsen, J.~Hald, and J.~C. Peterson, \enquote{{Saturated absorption in
			acetylene and hydrogen cyanide in hollow-core photonic bandgap fibers},}
	{\protect\JournalTitle{Optics Express}} \textbf{13}, 10475 (2005).
	
	\bibitem{Cizek2022}
	M.~{\v{C}}{\'{i}}{\v{z}}ek, L.~Pravdov{\'{a}}, T.~{Minh Pham},
	A.~Le{\v{s}}und{\'{a}}k, J.~Hrabina, J.~Lazar, T.~Pronebner, E.~Aeikens,
	J.~Premper, O.~Havli{\v{s}}, R.~Velc, V.~Smotlacha, L.~Altmannov{\'{a}},
	T.~Schumm, J.~Vojt{\v{e}}ch, A.~Niessner, and O.~{\v{C}}{\'{i}}p,
	\enquote{{Coherent fibre link for synchronization of delocalized atomic
			clocks},} {\protect\JournalTitle{Optics Express}} \textbf{30}, 5450 (2022).
	
	\bibitem{panek2012}
	P.~P{\'a}nek and A.~Kuna, \enquote{Time scales comparisons using simultaneous
		measurements in three frequency channels,} in \emph{Proceedings of the 44th
		Annual Precise Time and Time Interval Systems and Applications Meeting,}
	(2012), pp. 301--310.
	
	\bibitem{Cip2018}
	O.~{\v{C}}{\'{i}}p, A.~Le{\v{s}}und{\'{a}}k, T.~M. Pham, V.~Hucl,
	M.~{\v{C}}{\'{i}}{\v{z}}ek, J.~Hrabina, {\v{S}}.~{\v{R}}e{\v{r}}ucha,
	J.~Lazar, P.~Ob{\v{s}}il, R.~Filip, and L.~Slodi{\v{c}}ka, \enquote{{The
			compact setup for laser cooling and high-resolution spectroscopy with cold
			$^{40}$Ca$^+$ ions},} {\protect\JournalTitle{2018 European Frequency and Time
			Forum, EFTF 2018}} pp. 392--394 (2018).
	
	\bibitem{Riehle_2018}
	F.~Riehle, P.~Gill, F.~Arias, and L.~Robertsson, \enquote{The {CIPM} list of
		recommended frequency standard values: guidelines and procedures,}
	{\protect\JournalTitle{Metrologia}} \textbf{55}, 188--200 (2018).
	
	\bibitem{Schodel2021}
	R.~Sch{\"{o}}del, A.~Yacoot, and A.~Lewis, \enquote{The new mise en pratique
		for the metre - a review of approaches for the practical realization of
		traceable length metrology from $10^{-11}$m to $10^{-13}$m,}
	{\protect\JournalTitle{Metrologia}} \textbf{58} (2021).
	
\end{thebibliography}



\end{document}